\documentclass{elsart}

\usepackage{amssymb}
\usepackage{graphicx}
\begin{document}

\begin{frontmatter}

\title{What can we learn from the Interacting Boson Model in the limit of 
large boson numbers?  }
\author[1]{Dennis Bonatsos},
\author[2]{E. A. McCutchan},
\author[3]{R. F. Casten}

\address[1]{Institute of Nuclear Physics, N.C.S.R. ``Demokritos'', GR-15310 Aghia Paraskevi, Attiki, Greece}


\address[2]{Physics Division, Argonne National Laboratory, Argonne, Illinois 60439, USA}


\address[3]{Wright Nuclear Structure Laboratory, Yale University, New Haven, Connecticut 06520-8124, USA}


\begin{abstract}

Over the years, studies of collective properties of medium and heavy mass nuclei 
in the framework of the Interacting Boson Approximation (IBA) model have focused on finite boson numbers, 
corresponding to valence nucleon pairs in specific nuclei. Attention to large boson numbers 
has been motivated by the study of shape/phase transitions from one limiting symmetry of IBA to another,
which become sharper in the large boson number limit, revealing in parallel regularities previously unnoticed, 
although they survive to a large extent for finite boson numbers as well.  
Several of these regularities will be discussed. It will be shown that in all of the three 
limiting symmetries of the IBA [U(5), SU(3), and O(6)], energies of $0^+$ states grow linearly 
with their ordinal number. Furthermore, it will be proved that the narrow transition region separating 
the symmetry triangle of the IBA into a spherical and a deformed region is described quite well 
by the degeneracies $E(0_2^+)=E(6_1^+)$, $E(0_3^+)=E(10_1^+)$, $E(0_4^+)=E(14_1^+)$, the energy ratio 
$E(6_1^+) /E(0_2^+)$ turning out to be a simple,
empirical, easy-to-measure effective order parameter, distinguishing between first- and second-order
transitions. The energies of $0^+$ states near the point of the first order shape/phase transition 
between U(5) and SU(3) 
will be shown to grow as n(n+3), where $n$ is their ordinal number, in agreement with the rule dictated by 
the relevant critical point symmetries studied in the framework of special solutions of the Bohr Hamiltonian. The underlying dynamical and quasi-dynamical symmetries are also discussed.

\end{abstract}

\end{frontmatter}



Collective phenomena in atomic nuclei are described in terms of two complementary models,
the algebraic Interacting Boson Approximation (IBA) model \cite{IA}, and the geometrical collective model
\cite{Bohr,BM}. In the former, $s$ and $d$ bosons (bosons of angular momentum 0 and 2 respectively)
are used, while in the latter the collective variables $\beta$ (the ellipsoidal deformation)
and $\gamma$ (a measure of axial asymmetry) occur. The characteristic nuclear shapes occuring in the IBA 
are depicted at the vertices 
of the symmetry triangle \cite{Casten} of the model (Fig.~1), labeled by their underlying 
dynamical symmetries, which are
 i) U(5), corresponding to near-spherical (vibrational) nuclei, 
 ii) SU(3), representing axially symmetric prolate deformed (rotational) nuclei, and
 iii) O(6), describing nuclei soft with respect to axial asymmetry ($\gamma$-unstable).
   
\begin{figure} 
\center{{\includegraphics[height=60mm]{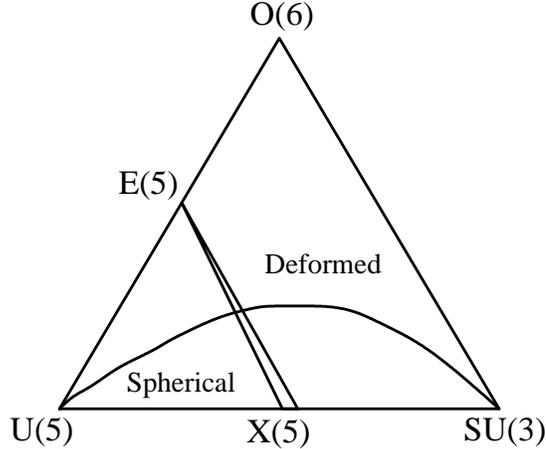}}}
\caption{IBA symmetry triangle with the three dynamical symmetries. The critical point models E(5) and X(5) are placed
close to the phase transition region (slanted
lines). The solid curve indicates the Alhassid-Whelan arc of regularity. Adopted from Ref. \cite{PRL}.}
\end{figure}
 
Shape/phase transitions from one nuclear shape to another were first discussed in the context of the IBA
in Ref. \cite{Feng}, applying catastrophe theory to the energy functional \cite{IZC} obtained 
in the classical limit of the IBA Hamiltonian through the use of the coherent state formalism
\cite{GK,DSI}.  A first order phase transition (in the Ehrenfest classification) has been found 
between the limiting symmetries U(5) and SU(3), while a second order phase transition has been located  
between U(5) and O(6). The spherical and deformed phases are separated by a narrow shape coexistence 
region \cite{IZC} (also shown in Fig.~1), shrinking into the point of second order phase transition as the U(5)-O(6) line is approached. 

More recently, shape/phase transitions have been considered also in the framework of the geometrical collective 
model, resulting in the introduction of the critical point symmetries E(5) \cite{IacE5} and X(5) \cite{IacX5}. 
E(5), which corresponds to the second order transition between U(5) and O(6), is a special solution of the 
Bohr Hamiltonian \cite{Bohr} using a potential $u(\beta)$ independent of $\gamma$ and having the shape of an infinite square well potential in $\beta$. X(5), which corresponds to the first order transition between U(5) and SU(3), is a special solution of the Bohr Hamiltonian using a potential of the form $u(\beta)+v(\gamma)$,
where $u(\beta)$ is the same as before, while $v(\gamma)$ is a steep harmonic oscillator centered around $\gamma=0$. E(5) and X(5) are also shown in Fig.~1, close to the 
points of the second and first order phase transitions of the IBA, respectively. 

The competition between regular behavior dictated by underlying symmetries and chaotic behavior
corresponding to lack of symmetries, has been studied throughout the symmetry triangle of Fig.~1. 
A highly regular region has been found along the U(5)-O(6) line, expected because of the underlying O(5) 
symmetry known to be preserved along this line \cite{Talmi}. Quite surprisingly, another regular region
\cite{AW}, connecting U(5) to SU(3) and called the Alhassid--Whelan arc of regularity,
has been found inside the triangle, as shown in Fig.~1. The symmetry underlying the arc is yet unknown. 

\begin{table}
\caption{Order $\nu$ for states with any $J$, and for the special case of $J^{\pi}=0^+$ states, in the geometrical models E(5), X(5), Z(5), Z(4), and X(3). $J$ is the spin of the level, $\tau=J/2$, and $n_{w}$ is the wobbling quantum number \cite{BM} which is zero for $0^+$ states.}
\renewcommand{\arraystretch}{2.0}

\center{
\begin{tabular}{lcc|lcc}
\hline

Model & $\nu$  & $\nu_{J=0^+}$ &
Model & $\nu$  & $\nu_{J=0^+}$ \\

\hline

E(5) & $\tau$ + $\frac{3}{2}$ & $\frac{3}{2}$ &   &     &   \\

\hline

X(5) & $\sqrt{\frac{J(J+1)}{3} + \frac{9}{4}}$ &  $\frac{3}{2}$ &
Z(5) & $\frac{\sqrt{J(J+4)+3n_w(2J-n_w)+9}}{2}$ &  $\frac{3}{2}$ \\

\hline 

X(3) & $\sqrt{\frac{J(J+1)}{3}+\frac{1}{4}}$ & $\frac{1}{2}$ & 
Z(4) & $\frac{\sqrt{J(J+4)+3n_w(2J-n_w)+4}}{2}$ &  1 \\
\hline

\end{tabular}
}
\end{table}

\begin{table}
\caption{(Left) Energies of $0^+$ states in the E(5), Z(5), and X(5)
models. Energies on the left are in units of $E$($2_1^+$) = 1.0,
while in the column Norm, in units $E$($0_2^+$) = 1.0. The
normalized results are identical for each of the models. The
column IBA-Norm gives the normalized $0^+$ energies for a large
$N_B$ IBA calculation near the critical point. (Middle) Same for the Z(4) model.
(Right) Same for the X(3) model. Adopted from Ref. \cite{PRL}.}

\center{
\begin{tabular}{c|c|c|c|c|c||c|c||c|c}
\hline

$0_i^+$ & E(5) & Z(5) & X(5) & Norm & IBA-Norm & Z(4) & Norm & X(3) & Norm \\
\hline
$0_1^+$ & 0 & 0 & 0 & 0  & 0 & 0 & 0 & 0 & 0 \\
$0_2^+$ &  3.03 &  3.91 &  5.65 & 1.0  & 1.0  &  2.95 & 1.0  &  2.87 & 1.0  \\
$0_3^+$ &  7.58 &  9.78 & 14.12 & 2.50 & 2.48 &  7.60 & 2.57 &  7.65 & 2.67 \\
$0_4^+$ & 13.64 & 17.61 & 25.41 & 4.50 & 4.62 & 13.93 & 4.71 & 14.34 & 5.00 \\
$0_5^+$ & 21.22 & 27.39 & 39.53 & 7.00 & 7.13 & 21.95 & 7.43 & 22.95 & 8.00 \\ 
$0_6^+$ & 30.31 & 39.12 & 56.47 &10.00 & 9.85 & 31.65 &10.72 & 33.47 &11.67 \\
\hline
\end{tabular}
}
\end{table}

$0^+$ states are particularly appropriate 
for the detection of underlying symmetries, because of the lack of centrifugal effects. 
As seen in Table~1, the critical point symmetries E(5) and X(5) mentioned above, as well as 
the Z(5) model \cite{Z5} [a solution of the Bohr Hamiltonian similar to X(5), using an infinite square well
in $\beta$, but with $v(\gamma)$ centered 
around $\pi/6$] possess as eigenfunctions  the Bessel functions $J_\nu$. While for $J \neq 0$ 
the order $\nu$ of the Bessel functions is different in each solution, for $J=0$ the same order 
is obtained in all three cases. As a result, $0^+$ states in these models look different if normalized 
to the energy of the $2_1^+$ state, but they become exactly identical if normalized 
to $0_2^+$, as seen in Table~2. Going further, one sees that in the latter normalization the energies 
of $0^+_n$ states, where $n$ is their ordinal number, follow the simple rule $n(n+3)$. 
This is due to the fact that the spectrum of the roots of the Bessel functions $J_\nu$
follows the $n(n+\nu+3/2)$ rule to a very good approximation for low $\nu$, being exact for $\nu=1/2$
\cite{PRL}. As a consequence, $0_n^+$ states in the Z(4) model \cite{Z4} [similar to Z(5), but with $\gamma$ 
fixed to $\pi/6$] follow the rule $n(n+2.5)$. Also, $0_n^+$ states in the X(3) model \cite{X3} [similar 
to X(5), but with $\gamma$ fixed to 0] follow the rule $n(n+2)$, as seen from Tables~1 and~2.    

Taking into account the second order Casimir operator of the E(5) algebra \cite{Z4,Barut}, the Euclidean algebra 
in 5 dimensions, one can see that the 
$0_n^+$ states in X(5) and Z(5) represent a case of a partial dynamical symmetry \cite{AL} of Type I
\cite{Leviatan}, a situation in which part of the states (the $0^+$ states in the present case) 
preserve the whole symmetry. 

What is a nontrivial result \cite{PRL}, is that an IBA calculation near the point of the first order phase 
transition leads to a spectrum of $0_n^+$ states also following the $n(n+3)$ rule, dictated by 
infinite well potentials used in the Bohr Hamiltonian utilizing 5 degrees of freedom (the collective 
variables $\beta$, $\gamma$, as well as the three Euler angles), as seen in Table~2. IBA calculations 
are performed using the usual IBA Hamiltonian \cite{Werner}, involving two parameters ($\zeta$, $\chi$). 
Large boson numbers can be reached using the recently developed IBAR code \cite{IBAR,Liz}.   

\begin{table}
\caption{$0^+$ bandheads in various $(\lambda, \mu)$ irreps of the SU(3) limit of IBA.
$N$ stands for the boson number, $N_B$.
Adopted from Ref. \cite{PRL}.}
\center{
\begin{tabular}{ l l l l  }
\hline 
irrep & $0^+$ & irrep & $0^+$ \\
\hline
(2N,0)    & 0              &           &                \\
(2N-4,2)  & 1              &           &                \\
(2N-8,4)  & (4N-6)/(2N-1)  & (2N-6,0)  & (4N-3)/(2N-1)  \\
(2N-12,6) & (6N-15)/(2N-1) & (2N-10,2) & (6N-10)/(2N-1) \\
(2N-16,8) & (8N-28)/(2N-1) & (2N-14,4) & (8N-21)/(2N-1) \\ 
\hline
\end{tabular}
}
\end{table}

\begin{table}
\caption{$0^+$ bandheads in various $(\sigma)$ irreps of the O(6) limit of IBA.
$N$ stands for the boson number, $N_B$. Adopted from Ref. \cite{PRL}.}
\bigskip
\center{
\begin{tabular}{ l l l  l l l  l l l l }
\hline 
irrep & $0^+$ & irrep & $0^+$ & irrep & $0^+$ & irrep & $0^+$ & irrep & $0^+$ \\
\hline
(N) & 0 & (N-2) & 1 & (N-4) & 2 & (N-6) & 3-3/N & (N-8) & 4-8/N \\ 
\hline
\end{tabular}
}
\end{table}

It should be noticed that the $n(n+3)$ behavior of the $0_n^+$ states of the IBA near the critical point 
of the first order phase transition is very different from the behavior of $0_n^+$ states obtained 
in the three dynamical symmetries of the model. Indeed, in the U(5) limit the energies of $0^+$ 
states increase linearly with the number of $d$ bosons, i.e. with the phonon number. Analytical results for 
$0^+$ bandheads in the SU(3) and O(6) limits of the IBA are shown in Tables~3 and~4. In both cases in the limit 
of large boson numbers a linear increase is obtained. We conclude that in all three dynamical symmetries 
of the IBA, $0_n^+$ bandheads in the large boson number limit increase linearly, $E=A n$. 

\begin{figure} 
\center{{\includegraphics[width=100mm]{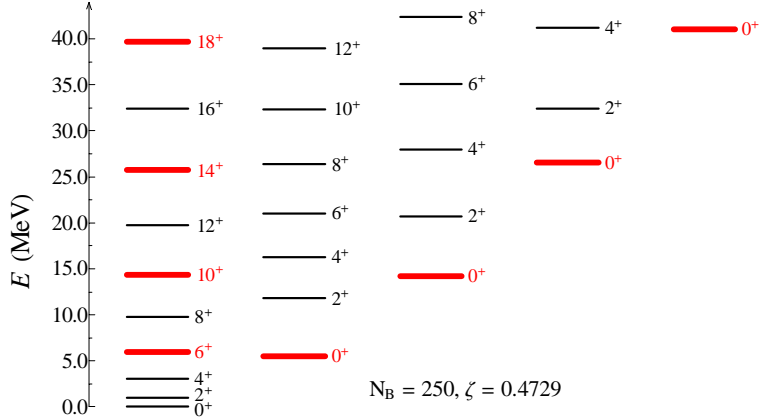}}}
\caption{Energies of low-lying states (normalized
to $E$($2_1^+$)=1) of the usual IBA Hamiltonian \cite{Werner} with
$\chi$=$-\sqrt{7}/2$, $\zeta$=0.4729, and $N_B$=250.
$\zeta$ was chosen to reproduce the approximate degeneracy of
$E$($0_2^+$) and $E$($6_1^+$). Adopted from Ref. \cite{PRL100}.}
\end{figure}

The regular behavior of $0^+$ states near the point of first order phase transition in the IBA 
invites a search for regularities of states with nonzero angular momenta. Indeed, as seen in Fig.~2
in an IBA calculation near the critical point, the $0^+$ bandheads are approximately degenerate with 
alternate levels of the ground state band with $J/2$ odd. 

\begin{figure} 
\center{\includegraphics[width=135mm]{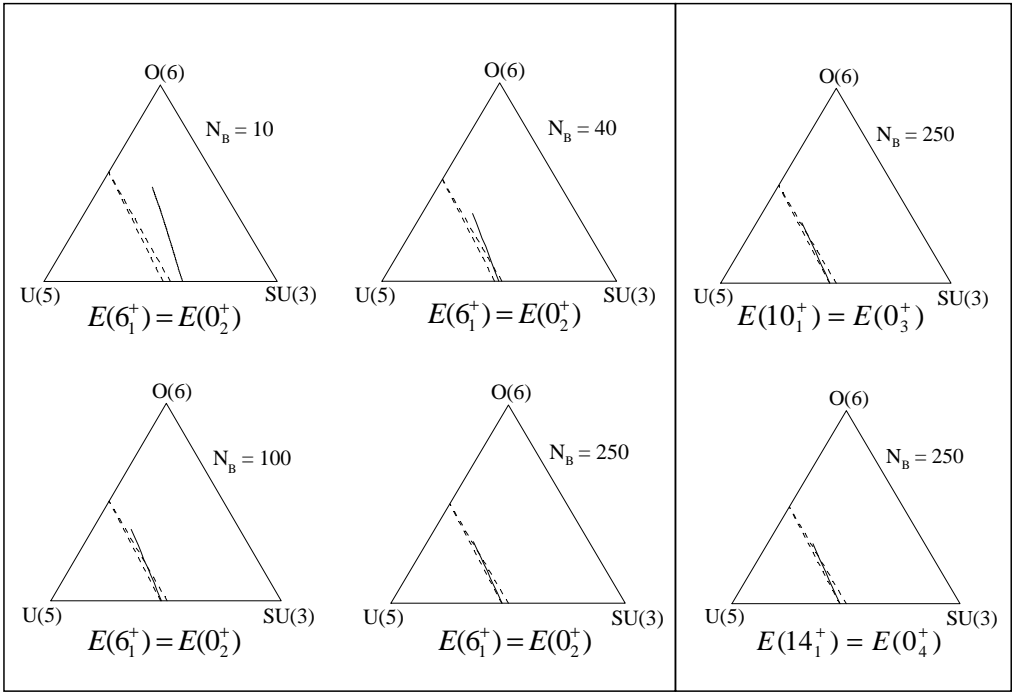}}
\caption{ (Left) Line of degeneracy between the
$0_2^+$ and $6_1^+$ levels for $N_B$ = 10, 40, 100,
and 250 in the IBA triangle. (Right) Line of degeneracy between
the $0_3^+$ and $10_1^+$ levels for $N_B$ = 250
(top) and between the $0_4^+$ and $14_1^+$ levels 
 for $N_B$ = 250 (bottom) in the IBA triangle. The dashed
lines denote the critical region in the IBA obtained in the large
$N_B$ limit from the intrinsic state formalism. Adopted from Ref. \cite{PRL100}.}
\end{figure}

Further investigation \cite{PRL100} of these degeneracies shows that the locus of the 
degeneracy $E(0_2^+)=E(6_1^+)$ [which is a hallmark of X(5)] in the IBA symmetry triangle 
is a straight line approaching the coexistence region in the limit of large boson numbers,
as seen in Fig.~3. Similar results are obtained for the degeneracies $E(0_3^+)=E(10_1^+)$ and $E(0_4^+)=E(14_1^+)$. One concludes that these degeneracies characterize the coexistence region 
until the U(5)-O(6) line is approached. 

\begin{figure} 
\center{\includegraphics[width=135mm]{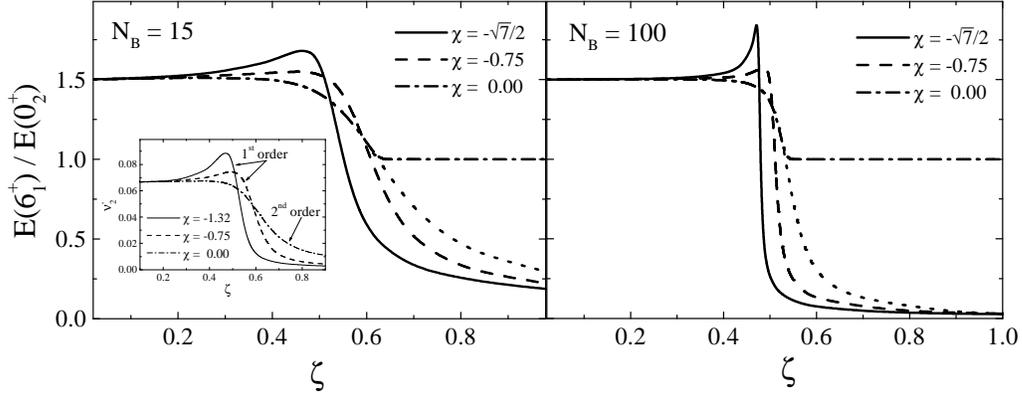}}
\caption{ The ratio $E$($6_1^+$)/$E$($0_2^+$) as a
function of $\zeta$ for three values of $\chi$ for (a) $N_B$ = 15
and (b) $N_B$ = 100. The inset to (a) shows the corresponding
behavior for $\nu_2^{\prime}$~\cite{IZ}. Adopted from Ref. \cite{PRL100}.}
\end{figure}

The ratio $E(6_1^+)/E(0_2^+)$ \cite{PRL100}, related to the first of the degeneracies 
mentioned above, turns out to be a simple empirical order parameter able to distinguish
between a first order and a second order phase transition. Indeed, as seen in Fig.~4, 
this ratio exhibits the same behavior as the order parameter $\nu'_2$ used in Ref. \cite{IZ}. 
A first order phase transition is seen for $\chi=-\sqrt{7}/2$, while a second order phase transition 
is seen for $\chi=0$. The ratio exhibits a sharp maximum just before the critical point in the first case
(the effect becoming stronger at larger boson numbers), while in the second case its behavior is smooth. 

\begin{figure} 
\center{\includegraphics[width=135mm]{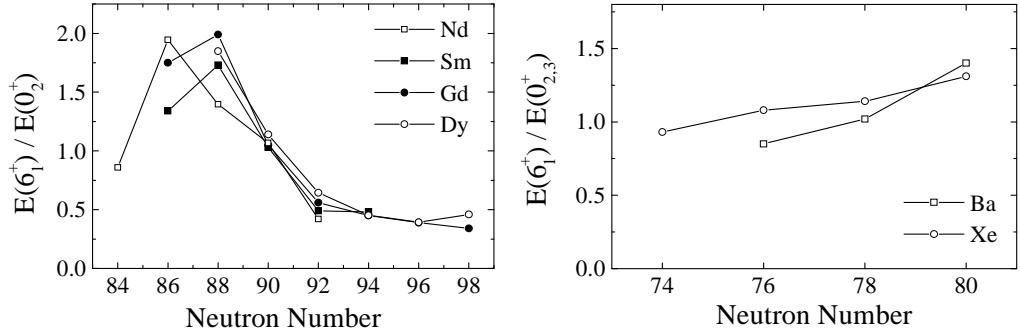}}
\caption{(a) Experimental $E$($6_1^+$)/$E$($0_2^+$)
ratio as a function of neutron number for the Nd, Sm, Gd, and Dy
isotopes. (b) Same for the Xe and Ba isotopes. For smaller neutron
numbers, the $0_3^+$ state was taken in the ratio if its $B$($E$2)
decay was consistent with the $\sigma$ = $N-2$ state. This
corresponds to $N$ = 74 in Xe and $N$ = 76,78 in Ba. Valence (hole) neutron number 
increases to the left. Adopted from Ref. \cite{PRL100}.}
\end{figure}

Experimental data around the $N=90$ isotones, the best empirical examples of X(5) 
\cite{CZX5,Kruecken,Tonev,Dewald,review},
do exhibit in Fig.~5(a) a clear maximum just before $N=90$, in agreement with the behavior 
expected for a first order phase transition. In contrast, experimental  
data around $^{134}$Ba, the best example of E(5) \cite{CZE5,review}, shows in Fig. 5(b) 
the smooth behavior expected for a second order transition.  

In conclusion, the large boson number limit of IBA reveals many regularities for $0^+$ states, 
the ones near the critical point of the first order phase transition being very close to the behavior 
obtained in critical point symmetries in the framework of the Bohr Hamiltonian. Degeneracies of $0^+$ 
states with states of nonzero angular momentum turn out to characterize the coexistence 
region separating the spherical phase from the deformed one, while ratios of energies of such pairs of states, 
like $E(6_1^+)/E(0_2^+)$, turn out to serve as order parameters able to distinguish between first order 
and second order phase transitions.  These degeneracies call for further investigations into 
finding the symmetries underlying them. The recent conjecture \cite{Macek} of a partial SU(3) dynamical symmetry underlying the Alhassid--Whelan arc of regularity is also receiving attention.  

Work supported in part U.S. DOE Grant No. DE-FG02-91ER-40609 and  under Contract DE-AC02-06CH11357.

\end{document}